\documentclass[twoside,12pt]{report}
\usepackage[T1]{fontenc}
\usepackage{times} 
\usepackage{geometry}
\usepackage{graphicx} 
\usepackage[table]{xcolor}
\usepackage{color}

\usepackage{natbib}
\usepackage{amsmath}
\usepackage{mathtools}
\usepackage{amsfonts}
\usepackage{float}
\usepackage{subcaption}
\usepackage{dsfont}
\long\def\TS2#1{{\color{blue}{[TS2: #1]}\color{black}}}

\newcommand{\argmin}{\mathop{\mathrm{argmin}}}

\title{Ranking probabilistic forecasting models with different loss functions}
\author{Tomasz Serafin\footnote{Department of Operations Research and Business Intelligence, Wrocław University of Science and Technology, 50-370 Wrocław, Poland; \textit{email:} t.serafin@pwr.edu.pl}, Bartosz Uniejewski$^1$ } 
\date{\today}

\begin{document}

\maketitle

\tableofcontents

\chapter{Introduction}

In the recent publication, \cite{mac:uni:wer:23} distinguish three main trends visible in the electricity price forecasting literature:  
\begin{enumerate}
    \item rising interest in probabilistic electricity price forecasts,
    \item departure from simple regression models towards more advanced forecasting models,
    \item evaluation of electricity price predictions in terms of economic measures.
\end{enumerate}

In this study we explore the crossover of trends $\#1$ and $\#3$ and investigate the relation of economic and statistical measures in the context of probabilistic electricity price forecasting. We introduce various metrics based on the well-recognized statistical measures (\cite{gne:raf:07}) for the evaluation of probabilistic forecast and rank the forecasting models according to their results. Then, forecasts from the best performing models are used in a real-life trading scenario involving the operation of a battery energy storage system (BESS) \cite{uni:24}), i.e. buying electricity in the hour with the lowest forecasted price and selling it in the hour with the highest forecasted price. We show that ranking the models according to the coverage of quantile forecasts in the trading hours outperforms all other approaches and produces highest profits-per-trade in the out-of-sample period. 

\chapter{Forecasting models}
Probabilistic electricity price forecasting gained interest of researchers after the Global Energy Forecasting Competition in 2014 (\cite{hon:pin:fan:etal:16}). Their advantage over most commonly used point forecasts (limited to a single value), includes information about the distribution of future electricity prices. This is typically represented as a collection of prediction intervals (PIs), in the form of 99 percentiles or a cumulative distribution function of price distribution (\cite{mac:uni:wer:23}). In the literature, two main approaches to generating probabilistic forecasts stand out: directly considering the distribution of electricity prices (assuming a certain distribution as Normal or Johnson's SU (\cite{mar:nar:wer:zie:23})) or deriving the approximate distribution based on point forecasts and the errors associated with them. In this paper, we follow \cite{uni:24} and consider nine different forecasting models, utilizing the latter of the approaches mentioned above. This chapter introduces all forecasting models used in the later part of this paper.

\section{Point forecasting model}
\label{sec:point_forecasts}
All probabilistic forecasting models considered here utilize point forecasts as the base for approximating the distribution of future electricity prices. Here, the parsimonious autoregressive model of \cite{zie:wer:18} is used for delivering electricity price point forecasts for day $d$ and hour $h$:
\begin{eqnarray}
	\label{eqn:expert}
	Y_{d,h} & = & \underbrace{\beta_1 Y_{d-1,h} + \beta_2 Y_{d-2,h} + \beta_3 Y_{d-7,h}}_{\mbox{\scriptsize autoregressive effects}} + \underbrace{\beta_4 Y_{d-1,24}}_{\mbox{\scriptsize end-of-day}} + \underbrace{\beta_5 Y_{d-1}^{max} + \beta_6 Y_{d-1}^{min}}_{\mbox{\scriptsize non-linear effects}} \nonumber \\
	&& + \underbrace{\beta_7 L_{d,h}}_{\mbox{\scriptsize load}} + \underbrace{\sum\nolimits_{j=1}^7 \beta_{h,j+7} D_{j}}_{\text{weekday dummies}} + \varepsilon_{d,h}, 
\end{eqnarray}
where $Y_{d-1,h}$, $Y_{d-2,h}$ and $Y_{d-7,h}$ represent the lagged prices at the same hour of the previous day, two days earlier, and one week earlier, respectively. $Y_{d-1,24}$ is the last known price at the time of forecasting, whereas $Y^{max}_{d-1}$ and $Y^{min}_{d-1}$ are the maximum and minimum prices of the previous day. $L_{d,h}$ is the day-ahead load forecast for a given hour of a day, finally $D_{1},..., D_{7}$ are weekday dummies, identifying the weekly seasonality. $\varepsilon_{d,h}$ is the noise term, assumed to be independent and identically distributed variables.

\section{Probabilistic forecasting models}
\label{sec:prob_forecasts}
All considered probabilistic forecasting models (\cite{uni:24}) build upon point forecasts from Section \ref{sec:point_forecasts}. This section provides descriptions and formulations of the aforementioned models. 

\subsection{Historical simulation (HS)}
Historical simulation is one of the simplest, yet robust benchmarks in probabilistic forecasting (\cite{mac:ser:uni:24}). It constructs probabilistic forecasts by considering both point forecasts and their associated errors. The electricity price quantile forecast ($\hat P^{q}_{d,h}$) is defined as:

\begin{equation}
	\hat P^{q}_{d,h} = \hat{P}_{d,h}+\gamma^{\text{HS}}_{q}, 
\end{equation}
where $\gamma_{q}$ is the $q$-th quantile of point forecast errors, $\hat{\varepsilon}_{d,h}$.

\subsection{Conformal Prediction (CP)}
\label{model:cp}
Conformal Prediction is yet another simple and robust method for deriving probabilistic forecasts from point forecast errors. It computes symmetrical prediction intervals (PI), centered around the point forecast, based on absolute point forecast errors within a calibration window \citep{sha:vov:08}. Here, the approach of \cite{kat:zie:21} is used, where
\begin{equation}
	\hat P^{q}_{d,h} = 	
	\begin{cases}
		\hat{P}_{d,h}-\gamma^{\text{CP}}_{q}, & \mbox{for } q<50\%,\\
		\hat{P}_{d,h}+\gamma^{\text{CP}}_{q}, & \mbox{for } q>50\%],
	\end{cases} 
\end{equation}
and the parameter $\gamma^{\text{CP}}_{q}$ is derived as the $|1-2q|$th quantile of $|\hat{\varepsilon}_{d,h}|$.

\subsection{Johnson distribution (ARX-J)}
This method assumes that electricity priced follow the \citet{joh:49} distribution (JSU), which has been shown in the literature (\citet{gia:bun:17}) to significantly outperform several other distributions including the normal (Gaussian). The model is defined as:

\begin{equation}
	\hat P^{q}_{d,h} = \hat{P}_{d,h}+\gamma^{\text{J}}_{q}, 
\end{equation}
where $\hat{P}_{d,h}$ is the point forecast and $\gamma_{q}$ is the $q$ quantile of the Johnson's distribution of the point forecast errors ($\hat{\varepsilon}_{d,h}$) with parameters estimated in the probabilistic calibration period.

\subsection{Quantile Regression (QR) Averaging}
 Introduced by \citet{now:wer:15}, quantile regression averaging is widely used in the electricity price forecasting literature \citep{jan:mic:20, kat:zie:21,mac:ser:uni:24}. QR Averaging utilizes quantile regression \citep{koe:05} in such a setup that point forecasts of the dependent variable, i.e. ${\hat P}^{(i)}_{d,h}, i=1,...,n$, are used as the regressors. In order to obtain model parameter estimates, the sum of the so-called pinball scores is minimized \citep{gne:raf:07}: 

\begin{equation}\label{eq:qr:loss}
	\hat{\boldsymbol\beta}_q
	= \underset{\boldsymbol\beta_{q}}\argmin \Big\{ \textstyle\sum_{d,h} \underbrace{\left({q}-{1}_{{P}_{d,h}< {X_{d,h}}\boldsymbol\beta_{q}}\right)\left({P}_{d,h} - X_{d,h} \boldsymbol\beta_q\right)}_{\mbox{\scriptsize \emph{pinball score}}} \Big\}.
\end{equation}

Despite its superb performance in GEFCom2014, the method exhibits performance decrease with the increasing number of regressors exceeding more than few \citet{mar:uni:wer:20}. In addition, the obtained interval predictions tend to be too narrow, making the electricity price forecast less reliable. To address these problems, \citet{wan:etal:19} proposed adding a constraint to the model so that the parameters are non-negative and sum to one. More recently, \citet{uni:wer:21} suggested using $\ell^1$ regularization to correctly select the inputs for QRA. While this solution yields significantly lower pinball loss, it does not completely solve the problem of too narrow prediction intervals.

\subsection{Smoothing Quantile Regression (SQR) Averaging}
\label{sssec:SQRA}
In a recent article, \citet{fer:gue:hor:21}, propose the modification of the standard quantile regression by smoothing the objective function using kernel estimation. Authors show that the proposed solution yields a lower mean squared error compared to the standard QR estimator and additionally possesses desirable theoretical properties. Recently, \cite{uni:24} proposed to use the smoothed QR to average a pool of point forecasts, creating a novel technique to obtain probabilistic forecast directly from point predictions. The estimator of model parameters $\boldsymbol\beta_{q}$ takes following form:
\begin{equation}\label{eq:sqr:loss}
	\hat{\boldsymbol\beta}_q
	= \underset{\boldsymbol\beta_{q}}\argmin \Big\{ \textstyle\sum_{d,h}  H \times \phi\left(\frac{{P}_{d,h} - X_{d,h} \boldsymbol\beta_q}{H}\right) + \left(q - \Phi\left(\frac{{P}_{d,h} - X_{d,h} \boldsymbol\beta_q}{H}\right)\right) \left({P}_{d,h} - X_{d,h} \boldsymbol\beta_q \right)  \Big\},
\end{equation}
where $X_{d,h}$ is the matrix of $n$ point forecasts ${\hat P}^{(i)}_{d,h}$ for $i=1,\ldots, n$, $\Phi(\cdot)$ and $\phi(\cdot)$ are the cumulative distribution function (cdf) and the probability density function (pdf) of the standard normal distribution, respectively, while $H$ is the so-called bandwidth parameter.

\chapter{Evaluation measures}

The majority of forecasting literature focuses on the statistical evaluation of predictions using such measures as mean absolute error (MAE) and root mean squared error (RMSE) for point, and pinball score or continuous ranked probability score (CRPS), for probabilistic forecasts. However, the interest of researchers is gradually shifting towards the evaluation of electricity price predictions in terms of economic measures (\cite{mac:uni:wer:23}). Intuitive interpretation, relation to practical problems and real-world scenarios are only few advantages of this particular way of forecast evaluation. \cite{yar:pet:21} claim that the literature generally reveals a disagreement between traditional error metrics and economic measures of performance. Although some studies consider the economic value of forecasts (\cite{kat:zie:18,ser:wer:24,mac:ser:uni:24}), to our best knowledge no study investigated this issue for linking statistical and economic performance in the context of probabilistic electricity price forecasting.

\section{Statistical measures}
\label{sec:stat_measures}

In this paper we consider two main measures for the statistical evaluation of probabilistic forecasts: pinball loss and empirical coverage.

\subsection{Pinball loss}
Pinball loss is an asymmetric scoring function for evaluating quantile predictions (\cite{gne:raf:07,gru:etal:17}):
\begin{equation}\label{eq:pinball}
    \text{PL}_{q}= \left({q}-{1}_{{P}_{d,h}< \hat{P}^{q}_{d,h}}\right)\left({P}_{d,h} - \hat{P}^{q}_{d,h}\right),
\end{equation}
where $q$ is the quantile order, ${P}_{d,h}$ is the actual electricity price for day $d$ and hour $h$ and $\hat{P}^{q}_{d,h}$ is the electricity price quantile forecast of order $q$.

\subsection{Empirical coverage}

Probabilistic forecasts can be represented in the form of prediction intervals (of order $\alpha$) - these are either directly derived from the forecasting model (see Section \ref{model:cp}) or created by selecting a pair of quantiles, $[\hat{P}^{(1-\alpha)/2}_{d,h}, \hat{P}^{(1+\alpha)/2}_{d,h}]$ (\cite{mac:ser:uni:24}). The competing forecasting schemes are compared on the basis of their empirical coverage \citep{cha:93}. For each day and hour of the testing period, we calculate the 'hits' of the prediction intervals:
\begin{equation}
I^{\alpha}_{d,h} = 
    \begin{cases}
        1 & \text{if } P_{d,h} \in PI_{\alpha}\\
        0 & \text{if } P_{d,h} \not\in PI_{\alpha}\\
    \end{cases}
\end{equation}
Next, the average number of 'hits' is calculated, which represents an empirical coverage level. For a given hour $h$, it is computed as
\begin{equation}
Cov^{\alpha}_h= \frac{1}{N}\sum_{d} I^{\alpha}_{d,h},
\end{equation}
The average coverage over all hours could be also calculated as:
\begin{equation}
Cov_{\alpha}= \frac{1}{24}\sum_{h} Cov^{\alpha}_h.
\end{equation}

\section{Economic measures}

A \cite{yar:pet:21} argue, the value of forecasts in markets of a financial nature can be evaluated
using different trading strategies. In the electricity price forecasting literature, many researchers take the perspective of the battery energy storage system (BESS) manager and try to optimize the decisions of buying and selling electricity in market based on the generated forecasts (\cite{mac:ser:uni:24}). In this study we utilize the strategy introduced by \cite{uni:24}, that is introduced and described in detail below.

\subsection{Trading strategy}
\label{ssec:strategies}

The primary tool for the economic evaluation of probabilistic forecasts in this paper is the strategy proposed by \cite{uni:24}. It considers a 2.5 MWh battery with a round-trip efficiency of 90\%. The company aims to buy energy and charge the battery when the price is low, and discharge and sell energy when the price is high. In order to trade only in the day-ahead market, the company limits the volume of a single transaction (charge or discharge) to 1~MWh. The proposed strategy aims to leave 1 MWh of energy in the energy storage system at the end of each day. However, depending on the acceptance of submitted bid or offer the battery may be in the following states at the beginning of the next day:

\begin{itemize}\itemsep 0em
\item $B_d = 0$ - battery is empty and there is no more energy to sell. On day $d^*$, when $B_{d^*} = 0$, the company will additionally place a market order to buy energy and charge the battery.
\item $B_d = 1$ - a day starts with a half-full storage. In this state, the battery can be both charged and discharged.
\item $B_d = 2$ - the capacity limit has been reached and the battery cannot be charged at the beginning of day $d$. On day $d^*$, when $B_{d^*} = 2$, the company will discharge the battery and sell energy.
\end{itemize}

\subsubsection{Quantile-based trading strategies}
\label{sssec:Quantile-based:strategies}

\begin{figure}[tb]
	\centering 
	\includegraphics[width = 0.73 \linewidth]{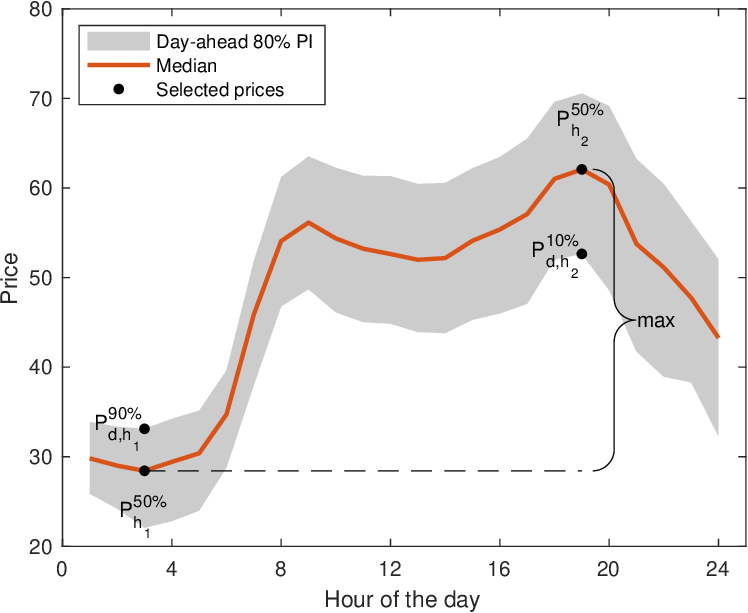}
	\caption{Illustration of the quantile-based bidding strategy for 17.1.2019 for and EPEX market. The orange line shows the median forecast ($\hat P^{50\%}_{d,h}$), the gray area represents the 80\% prediction interval computed one day earlier, and the filled circles at the PI bounds represent the offer price $\hat P^{10\%}_{d,h_2}$ and the bid price $\hat P^{90\%}_{d,h_1}$, see step (2) in Section \ref{sssec:Quantile-based:strategies}. Source \cite{uni:24}.}
	\label{fig:stategy}
\end{figure}

The proposed quantile-based trading strategies includes three steps that are performed each day. In step (1), two hours of the day - one with the lowest price $h_1$, and one with the highest price, $h_2$ are selected based on the median forecast ($\hat P^{50\%}$).In case the battery is fully discharged on day $d^*$ ($B_{d^*}=0$), an additional unlimited bid will be placed at hour $h^* < h_2$ in order to additionally charge the battery. Similarly, if $B_{d^*} = 2$, the company places an additional unlimited offer to sell energy before hour $h_2$. 

In step (2), having chosen the optimal hours for placing bids, the level of the bid and the offer must be determined. First, the trader arbitrarily chooses the level $\alpha$ of the prediction interval (PI). The bid price is equal to the upper bound of the PI at hour $h_1$ i.e., $\hat U_{d,h_1}^{\alpha} = \hat P_{d,h_1}^{\frac{1+\alpha}{2}}$, while the offer price is set equal to the lower limit of the PI at hour $h_2$ i.e., $\hat L_{d,h_2}^{\alpha} = \hat P_{d,h_1}^{\frac{1-\alpha}{2}}$. The prices $\hat U_{d,h_1}^{\alpha}$ and $\hat L_{d,h_1}^{\alpha}$ are marked with black dots in Figure \ref{fig:stategy}. 

Finally, in step (3), the profit for day $d$ is calculated depending on the acceptance of the bid and offer:
\begin{itemize}
	\item If both the bid and the offer are accepted, the daily profit is $0.9~P_{d,h_2} - \frac{1}{0.9}P_{d,h_1}$, and since the battery has been charged and discharged once, its state is unchanged ($B_{d+1} = B_{d}$).

	\item If only the offer is accepted, at hour $h_2$ the battery is discharged and 0.9 MWh of energy is sold on the day-ahead market. The daily profit is $0.9P_{d,h_2}$ and the battery state is reduced ($B_{d+1} = B_{d}-1$).

	\item Similarly, if only the bid is accepted, at hour $h_1$ the firm buys $\frac{1}{0.9}$ MWh of energy in the day-ahead market and charges the battery. The daily loss is $\frac{1}{0.9} P_{d,h_1}$, but the state of the battery is increased ($B_{d+1} = B_{d}+1$).
\end{itemize}

Additionally:
\begin{itemize}
	\item If $B_d = 0$, the trader additionally buys $\frac{1}{0.9}$ MWh of energy at hour $h^*$, so the daily profit decreases by $\frac{1}{0.9}P_{d,h^*}$ while the battery state increases.
	\item If $B_d = 2$, the battery owner additionally sells $0.9$ MWh of energy at hour $h^*$, so the daily profit increases by $0.9P_{d,h^*}$ while the battery state decreases.
\end{itemize}

\subsubsection{Unlimited-bids benchmark}
\label{sssec:benchmark:strategies}
A benchmark strategy can be proposed that does not require any knowledge of the price distribution. Trading is based on the hours with the lowest ($h_1$) and the highest ($h_2$) predicted prices (according to the point forecasts) each day. Unlimited (price taker) offers to sell energy at the highest price and to buy energy at the lowest price are then submitted.

\chapter{Methodology}

The main objective of this study is investigating the model discrimination ability of different statistical forecast evaluation methods in the context of probabilistic electricity price forecasting. More precisely, we utilize Pinball loss and empirical coverage (see Section \ref{sec:stat_measures}) to create several metrics for the evaluation of probabilistic forecasts from Section \ref{sec:prob_forecasts}. These metrics are later used to determine the ranking of forecasting models and their selection as inputs for the trading strategy. In the end, the overall model selection performance for each of the metrics is evaluated in the context of average profits obtained from trading activities. This section describes the above process in detail.

\section{Model selection}
In this study we consider probabilistic electricity price forecasts from 9 different models. The forecasts are assessed in terms of their statistical performance ($SP$) in the following ways (dubbed metrics):

\begin{itemize}
    \item By calculating the average daily pinball across all hours and quantiles for day $d$:
    $$ SP_{d}^{\text{Pinball all}} =  \frac{1}{24\cdot99} \sum_{q=0.01}^{0.99} \sum_{h=1}^{24} PL_{q,d,h} $$
    \item By calculating the average pinball of forecasts used for trading in hours $h_1$ and $h_2$ on day $d$:
    $$ SP_{d,\alpha}^{\text{Pinball buy/sell}} =  \frac{1}{2}\left(PL_{(1+\alpha)/2,d,h_1} + PL_{(1-\alpha)/2,d,h_2}\right)$$
    \item By calculating the pinball of forecast with the highest-price hour of day $d$, used for selling electricity in hour $h_2$:
    $$ SP_{d,\alpha}^{\text{Pinball sell}} =  PL_{(1-\alpha)/2,d,h_2} $$
    \item By calculating the pinball of forecast with the lowest-price hour of day $d$, used for buying electricity in hour $h_1$:
    $$ SP_{d,\alpha}^{\text{Pinball buy}} =  PL_{(1+\alpha)/2,d,h_1} $$
    \item By considering the average empirical coverage of PIs with nominal coverage $\alpha$ of day $d$:
    $$ SP_{d,\alpha}^{\text{Coverage all}} = \frac{1}{24} \sum_{h=1}^{24}I^{\alpha}_{d,h} $$
    \item By considering the empirical coverage of 'artificial' PIs based on quantiles used in the trading strategy:
    $$ SP_{d,\alpha}^{\text{Coverage hours}} = \begin{cases}
        1 & \text{if } P_{d,h_1} < \hat{P}^{(1+\alpha)/2}_{d,h_1}  \text{and } P_{d,h_2} > \hat{P}^{(1-\alpha)/2}_{d,h_2}\\
        0 & \text{if } \text{otherwise}\\
    \end{cases}  $$
\end{itemize}

For each day of the out-of-sample period, we calculate the statistical performance metrics for each of the considered models. Consecutively, a 30-day rolling window (see Figure \ref{fig:data}) is applied to calculate the most recent average statistical performance of the models according to each metric:

$$SP^{X} = \sum_{d=i}^{i+30} SP^{X}_{d},$$ 
where $X$ is a particular metric choice. Having calculated the average score, we can determine which forecasting model performed best over the last 30 days according to each metric. Next, we can simulate trading scenarios in which each metric selects forecasts from the best-performing model and use them for trading on the next day. This setup mimics a real-world decision-making problem of electricity trading companies having a number of different forecasting models at their disposal.

\section{Dataset}
\begin{figure}[tb]
	\centering 
	\includegraphics[width = 0.95 \linewidth]{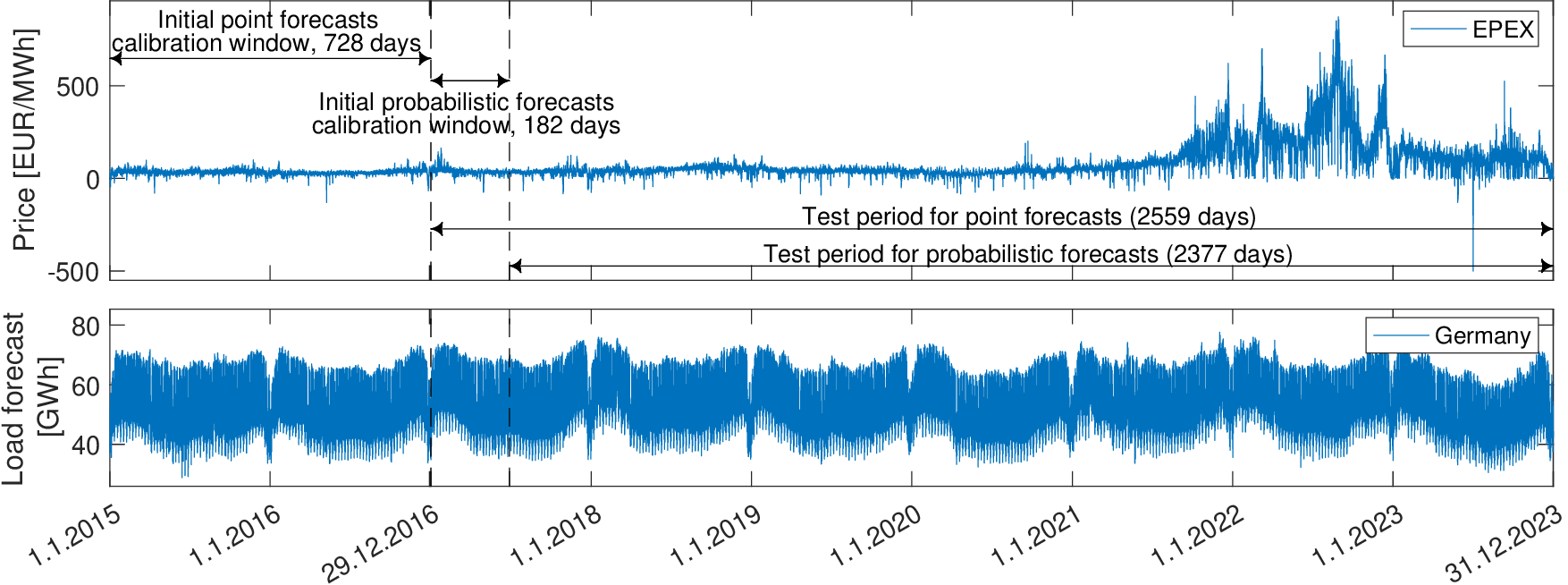}
	\caption{Dataset from the German day-ahead electricity market used in the study. Three rolling windows are considered: the first one for the calibration of point forecasting models (364 days), the second one for the calibration of probabilistic forecasting methods (182 days) and the third one for the calculation of statistical performance of probabilistic models for the considered metrics (30 days).}
	\label{fig:data}
\end{figure}

The methodology is tested on the dataset from the German day-ahead electricity market, spanning from 01.01.2015 to 31.12.2023 (see Figure \ref{fig:data}). Three rolling windows are considered: the first one for the calibration of point forecasting models (364 days), the second one for the calibration of probabilistic forecasting methods (182 days) and the third one for the calculation of statistical performance of probabilistic models for the considered metrics (182 days). The dataset provides a comprehensive test environment for the proposed methodology as it covers periods when prices exhibited varying behaviors - from low to high volatility regimes and periods of frequent positive and negative price spikes.  

\chapter{Results}

This section discusses the results from the evaluation of the proposed methodology. The trading strategy is parametrized by the prediction interval level $\alpha$ (see Section \ref{ssec:strategies}) - here we consider $\alpha = 50\%, \ldots, 98\%$. The economic value of forecasts is evaluated using the average profit per 1MWh of electricity (\cite{mac:ser:uni:24}). Total profit is not considered due to potential differences in trading volumes of the models. 

\begin{figure}[tb]
	\centering 
	\includegraphics[width = 0.95 \linewidth]{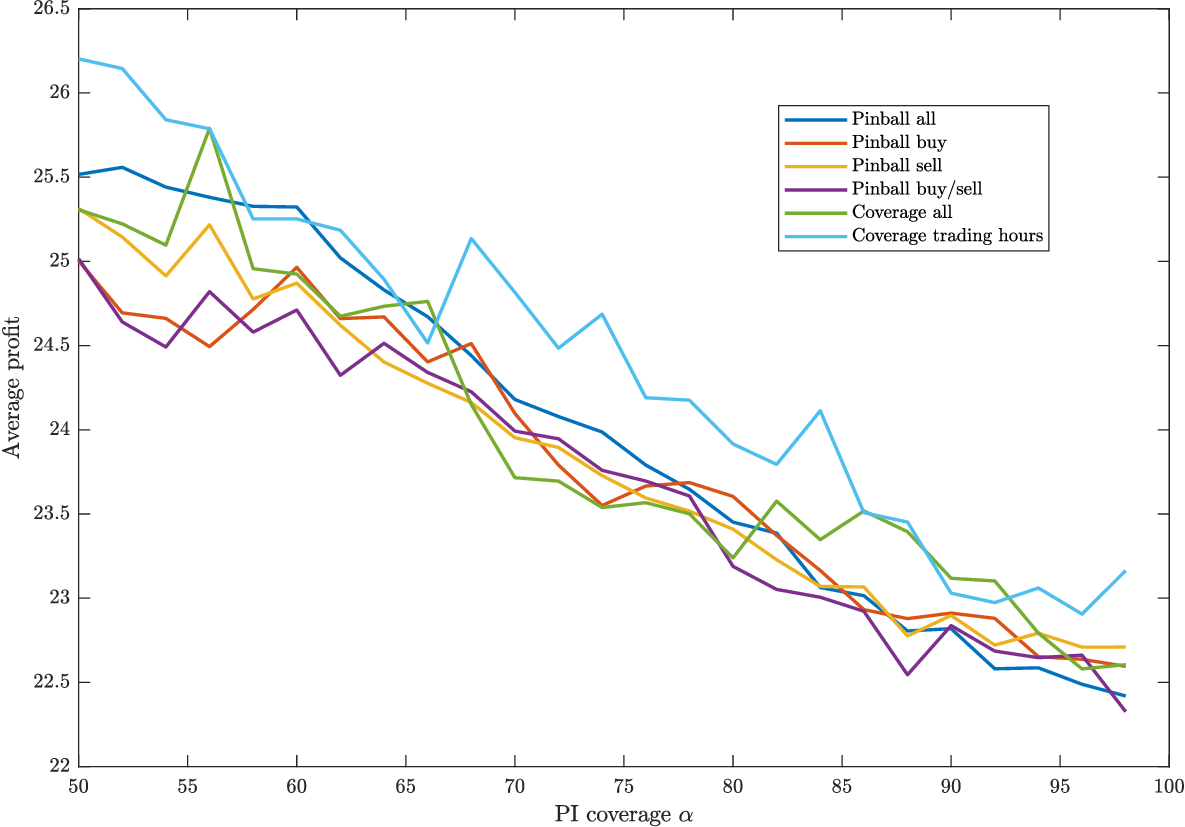}
	\caption{Comparison of average profits from the trading strategy for all considered statistical error measures ranking the forecasting models.}
	\label{fig:results}
\end{figure}

Figure \ref{fig:results} shows the average profits from trading based on the metric used to determine the best performing probabilistic forecasting model. Interestingly, there is a clear winner, outperforming all other metrics across almost all values of $\alpha$. Ranking the forecasting models according to the coverage of quantile forecasts used in the trading hours ($SP^{\text{Coverage hours}}$) exhibits a superior economic performance. While all Pinball-based metrics exhibit similar performance across the majority of $\alpha$ parameter values, ranking the model according to the average pinball loss across all hours and quantiles ($SP^{\text{Pinball all}}$) is the top performer, especially showing strength for $50\% < \alpha < 70\%$. Considering the avarage coverage across all hours ($SP^{\text{Coverage all}}$) performs well but shows greater variability in the performance.

\chapter{Conclusions}

In this study, we introduced various statistical performance metrics, based on the pinball loss and the empirical coverage, for the ranking of probabilistic forecasting models. We tested the ability of the proposed metrics to determine the top performing forecasting model and investigated the use of which metric corresponds to the highest average per-trade profit in the out-of-sample period. Our findings show that for the considered trading strategy,  ranking the forecasting models according to the coverage of quantile forecasts used in the trading hours ($SP^{\text{Coverage hours}}$) exhibits a superior economic performance. 

\subsubsection{Acknowledgments}
This work was partially supported by the Ministry of Science and Higher Education (MNiSW, Poland) through Diamond Grant No.\ 0009/DIA/2020/49 (to T.S.) and by the National Science Center (NCN, Poland) through grant no. 2023/49/N/HS4/02741 (to B.U.).

\bibliographystyle{elsarticle-harv}
\bibliography{main}

\begin{thebibliography}{20}
\expandafter\ifx\csname natexlab\endcsname\relax\def\natexlab#1{#1}\fi
\providecommand{\url}[1]{\texttt{#1}}
\providecommand{\href}[2]{#2}
\providecommand{\path}[1]{#1}
\providecommand{\DOIprefix}{doi:}
\providecommand{\ArXivprefix}{arXiv:}
\providecommand{\URLprefix}{URL: }
\providecommand{\Pubmedprefix}{pmid:}
\providecommand{\doi}[1]{\href{http://dx.doi.org/#1}{\path{#1}}}
\providecommand{\Pubmed}[1]{\href{pmid:#1}{\path{#1}}}
\providecommand{\bibinfo}[2]{#2}
\ifx\xfnm\relax \def\xfnm[#1]{\unskip,\space#1}\fi
\bibitem[{Chatfield(1993)}]{cha:93}
\bibinfo{author}{Chatfield, C.}, \bibinfo{year}{1993}.
\newblock \bibinfo{title}{Calculating interval forecasts}.
\newblock \bibinfo{journal}{Journal of Business \& Economic Statistics} \bibinfo{volume}{11}, \bibinfo{pages}{121--135}.
\bibitem[{Fernandes et~al.(2021)Fernandes, Guerre and Horta}]{fer:gue:hor:21}
\bibinfo{author}{Fernandes, M.}, \bibinfo{author}{Guerre, E.}, \bibinfo{author}{Horta, E.}, \bibinfo{year}{2021}.
\newblock \bibinfo{title}{Smoothing quantile regressions}.
\newblock \bibinfo{journal}{Journal of Business \& Economic Statistics} \bibinfo{volume}{39}, \bibinfo{pages}{338--357}.
\bibitem[{Gneiting and Raftery(2007)}]{gne:raf:07}
\bibinfo{author}{Gneiting, T.}, \bibinfo{author}{Raftery, A.}, \bibinfo{year}{2007}.
\newblock \bibinfo{title}{Strictly proper scoring rules, prediction, and estimation}.
\newblock \bibinfo{journal}{Journal of the American Statistical Association} \bibinfo{volume}{102}, \bibinfo{pages}{359--378}.
\bibitem[{Grushka-Cockayne et~al.(2017)Grushka-Cockayne, Lichtendahl~Jr, Jose and Winkler}]{gru:etal:17}
\bibinfo{author}{Grushka-Cockayne, Y.}, \bibinfo{author}{Lichtendahl~Jr, K.C.}, \bibinfo{author}{Jose, V.R.R.}, \bibinfo{author}{Winkler, R.L.}, \bibinfo{year}{2017}.
\newblock \bibinfo{title}{Quantile evaluation, sensitivity to bracketing, and sharing business payoffs}.
\newblock \bibinfo{journal}{Operations Research} \bibinfo{volume}{65}, \bibinfo{pages}{712--728}.
\bibitem[{Hong et~al.(2016)Hong, Pinson, Fan, Zareipour, Troccoli and Hyndman}]{hon:pin:fan:etal:16}
\bibinfo{author}{Hong, T.}, \bibinfo{author}{Pinson, P.}, \bibinfo{author}{Fan, S.}, \bibinfo{author}{Zareipour, H.}, \bibinfo{author}{Troccoli, A.}, \bibinfo{author}{Hyndman, R.J.}, \bibinfo{year}{2016}.
\newblock \bibinfo{title}{Probabilistic energy forecasting: {G}lobal {E}nergy {F}orecasting {C}ompetition 2014 and beyond}.
\newblock \bibinfo{journal}{International Journal of Forecasting} \bibinfo{volume}{32}, \bibinfo{pages}{896--913}.
\bibitem[{Johnson(1949)}]{joh:49}
\bibinfo{author}{Johnson, N.L.}, \bibinfo{year}{1949}.
\newblock \bibinfo{title}{{Systems of frequency curves generated by methods of translation}}.
\newblock \bibinfo{journal}{Biometrika} \bibinfo{volume}{36}, \bibinfo{pages}{149--176}.
\bibitem[{Kath and Ziel(2018)}]{kat:zie:18}
\bibinfo{author}{Kath, C.}, \bibinfo{author}{Ziel, F.}, \bibinfo{year}{2018}.
\newblock \bibinfo{title}{The value of forecasts: Quantifying the economic gains of accurate quarter-hourly electricity price forecasts}.
\newblock \bibinfo{journal}{Energy Economics} \bibinfo{volume}{76}, \bibinfo{pages}{411--423}.
\bibitem[{Kath and Ziel(2021)}]{kat:zie:21}
\bibinfo{author}{Kath, C.}, \bibinfo{author}{Ziel, F.}, \bibinfo{year}{2021}.
\newblock \bibinfo{title}{Conformal prediction interval estimation and applications to day-ahead and intraday power markets}.
\newblock \bibinfo{journal}{International Journal of Forecasting} \bibinfo{volume}{37}, \bibinfo{pages}{777--799}.
\bibitem[{Koenker(2005)}]{koe:05}
\bibinfo{author}{Koenker, R.W.}, \bibinfo{year}{2005}.
\newblock \bibinfo{title}{Quantile Regression}.
\newblock \bibinfo{publisher}{Cambridge University Press}.
\bibitem[{Maciejowska et~al.(2024)Maciejowska, Serafin and Uniejewski}]{mac:ser:uni:24}
\bibinfo{author}{Maciejowska, K.}, \bibinfo{author}{Serafin, T.}, \bibinfo{author}{Uniejewski, B.}, \bibinfo{year}{2024}.
\newblock \bibinfo{title}{Probabilistic forecasting with a hybrid factor-qra approach: Application to electricity trading}.
\newblock \bibinfo{journal}{Electric Power Systems Research} \bibinfo{volume}{234}, \bibinfo{pages}{110541}.
\newblock \DOIprefix\doi{https://doi.org/10.1016/j.epsr.2024.110541}.
\bibitem[{Maciejowska et~al.(2023)Maciejowska, Uniejewski and Weron}]{mac:uni:wer:23}
\bibinfo{author}{Maciejowska, K.}, \bibinfo{author}{Uniejewski, B.}, \bibinfo{author}{Weron, R.}, \bibinfo{year}{2023}.
\newblock \bibinfo{title}{Forecasting electricity prices}, in: \bibinfo{booktitle}{Oxford Research Encyclopedia of Economics and Finance}. \bibinfo{publisher}{Oxford University Press}.
\newblock \DOIprefix\doi{10.1093/acrefore/9780190625979.013.667}.
\bibitem[{Marcjasz et~al.(2023)Marcjasz, Narajewski, Weron and Ziel}]{mar:nar:wer:zie:23}
\bibinfo{author}{Marcjasz, G.}, \bibinfo{author}{Narajewski, M.}, \bibinfo{author}{Weron, R.}, \bibinfo{author}{Ziel, F.}, \bibinfo{year}{2023}.
\newblock \bibinfo{title}{Distributional neural networks for electricity price forecasting}.
\newblock \bibinfo{journal}{Energy Economics} \bibinfo{volume}{125}, \bibinfo{pages}{106843}.
\bibitem[{Marcjasz et~al.(2020)Marcjasz, Uniejewski and Weron}]{mar:uni:wer:20}
\bibinfo{author}{Marcjasz, G.}, \bibinfo{author}{Uniejewski, B.}, \bibinfo{author}{Weron, R.}, \bibinfo{year}{2020}.
\newblock \bibinfo{title}{Probabilistic electricity price forecasting with {NARX} networks: Combine point or probabilistic forecasts?}
\newblock \bibinfo{journal}{International Journal of Forecasting} \bibinfo{volume}{36}, \bibinfo{pages}{466--479}.
\bibitem[{Nowotarski and Weron(2015)}]{now:wer:15}
\bibinfo{author}{Nowotarski, J.}, \bibinfo{author}{Weron, R.}, \bibinfo{year}{2015}.
\newblock \bibinfo{title}{Computing electricity spot price prediction intervals using quantile regression and forecast averaging}.
\newblock \bibinfo{journal}{Computational Statistics} \bibinfo{volume}{30}, \bibinfo{pages}{791--803}.
\bibitem[{Serafin and Weron(2024)}]{ser:wer:24}
\bibinfo{author}{Serafin, T.}, \bibinfo{author}{Weron, R.}, \bibinfo{year}{2024}.
\newblock \bibinfo{title}{Loss functions in regression models: Impact on profits and risk in day-ahead electricity trading}.
\newblock \bibinfo{journal}{WORking papers in Management Science (WORMS), WORMS/24/03,} \bibinfo{volume}{Department of Operations Research and Business Intelligence, Wroclaw University of Science and Technology.}
\bibitem[{Uniejewski(2024)}]{uni:24}
\bibinfo{author}{Uniejewski, B.}, \bibinfo{year}{2024}.
\newblock \bibinfo{title}{Smoothing quantile regression averaging: A new approach to probabilistic forecasting of electricity prices}.
\newblock \URLprefix \url{https://arxiv.org/abs/2302.00411}, \href{http://arxiv.org/abs/2302.00411}{{\tt arXiv:2302.00411}}.
\bibitem[{Uniejewski and Weron(2021)}]{uni:wer:21}
\bibinfo{author}{Uniejewski, B.}, \bibinfo{author}{Weron, R.}, \bibinfo{year}{2021}.
\newblock \bibinfo{title}{Regularized quantile regression averaging for probabilistic electricity price forecasting}.
\newblock \bibinfo{journal}{Energy Economics} \bibinfo{volume}{95}, \bibinfo{pages}{105121}.
\bibitem[{Wang et~al.(2019)Wang, Zhang, Tan, Hong, Kirschen and Kang}]{wan:etal:19}
\bibinfo{author}{Wang, Y.}, \bibinfo{author}{Zhang, N.}, \bibinfo{author}{Tan, Y.}, \bibinfo{author}{Hong, T.}, \bibinfo{author}{Kirschen, D.}, \bibinfo{author}{Kang, C.}, \bibinfo{year}{2019}.
\newblock \bibinfo{title}{Combining probabilistic load forecasts}.
\newblock \bibinfo{journal}{IEEE Transactions on Smart Grid} \bibinfo{volume}{10}, \bibinfo{pages}{3664--3674}.
\bibitem[{Yardley and Petropoulos(2021)}]{yar:pet:21}
\bibinfo{author}{Yardley, E.}, \bibinfo{author}{Petropoulos, F.}, \bibinfo{year}{2021}.
\newblock \bibinfo{title}{Beyond error measures to the utility and cost of the forecasts}.
\newblock \bibinfo{journal}{Foresight} \bibinfo{volume}{Q4}, \bibinfo{pages}{36--45}.
\bibitem[{Ziel and Weron(2018)}]{zie:wer:18}
\bibinfo{author}{Ziel, F.}, \bibinfo{author}{Weron, R.}, \bibinfo{year}{2018}.
\newblock \bibinfo{title}{Day-ahead electricity price forecasting with high-dimensional structures: {U}nivariate vs. multivariate modeling frameworks}.
\newblock \bibinfo{journal}{Energy Economics} \bibinfo{volume}{70}, \bibinfo{pages}{396--420}.

\end{thebibliography}
\end{document}